\documentclass[sn-nature]{sn-jnl}

\usepackage{graphicx}%
\usepackage{multirow}%
\usepackage{amsmath,amssymb,amsfonts}%
\usepackage{amsthm}%
\usepackage{mathrsfs}%
\usepackage[title]{appendix}%
\usepackage{xcolor}%
\usepackage{textcomp}%
\usepackage{manyfoot}%
\usepackage{booktabs}%
\usepackage{algorithm}%
\usepackage{algorithmicx}%
\usepackage{algpseudocode}%
\usepackage{listings}%
\usepackage{mymacro}
\usepackage{bm}
\usepackage{comment}

\begin{document}

\title[Article Title]{Enhancing Intrinsic Quality Factors Approaching 10 Million in Superconducting Planar Resonators via Spiral Geometry}


\author*[1]{\fnm{Yusuke} \sur{Tominaga}}\email{yusuke.tominaga@riken.jp}

\author[1,2]{\fnm{Shotaro} \sur{Shirai}}\email{shotaro.shirai@riken.jp}

\author[3]{\fnm{Yuji} \sur{Hishida}}\email{yujihishida@nict.go.jp}

\author[3]{\fnm{Hirotaka} \sur{Terai}}\email{terai@nict.go.jp}


\author*[1,2,4]{\fnm{Atsushi} \sur{Noguchi}}\email{u-atsushi@g.ecc.u-tokyo.ac.jp}

\affil[1]{\orgname{RIKEN Center for Quantum Computing},  \orgaddress{\city{Wako}, \postcode{351-0198}, \state{Saitama}, \country{Japan}}}

\affil[2]{\orgdiv{Komaba Institute for Science (KIS)}, \orgname{The University of Tokyo}, \orgaddress{ \city{Meguro-ku}, \postcode{153-8902}, \state{Tokyo}, \country{Japan}}}

\affil[3]{\orgdiv{Advanced ICT Research Institute}, \orgname{National Institute of Information
and Communications Technology (NICT)}, \orgaddress{ \city{Kobe}, \postcode{651-2492}, \state{Hyogo}, \country{Japan}}}

\affil[4]{\orgdiv{Inamori Research Institute for Science (InaRIS)}, \orgname{Organization}, \orgaddress{ \city{Kyoto-shi}, \postcode{600-8411}, \state{Kyoto}, \country{Japan}}}


\abstract{This study investigates the use of spiral geometry in superconducting resonators to achieve high intrinsic quality factors, crucial for applications in quantum computation and quantum sensing. We fabricated Archimedean Spiral Resonators (ASRs) using domain-matched epitaxially grown titanium nitride (TiN) on silicon wafers, achieving intrinsic quality factors of $Q_\mathrm{i} = (9.6 \pm 1.5) \times 10^6$ at the single-photon level and $Q_\mathrm{i} = (9.91 \pm 0.39) \times 10^7$ at high power, significantly outperforming traditional coplanar waveguide (CPW) resonators.

We conducted a comprehensive numerical analysis using COMSOL to calculate surface participation ratios (PRs) at critical interfaces: metal-air, metal-substrate, and substrate-air. Our findings reveal that ASRs have lower PRs than CPWs, explaining their superior quality factors and reduced coupling to two-level systems (TLSs).}




\maketitle

\section{Introduction}\label{sec1}
Superconducting resonators are fundamental components in cutting-edge quantum technologies, playing essential roles in quantum sensing and quantum computation. They function as readout devices for superconducting quantum bits \cite{blais2004cavity, wallraff2004strong}, facilitate qubit interconnections \cite{sillanpaa2007coherent}, serve as quantum limited amplifiers \cite{bergeal2010phase}, and enable single-photon detection \cite{day2003broadband}.
Additionally, these devices are the key to storing and manipulating quantum information \cite{hofheinz2009synthesizing} exemplified by bosonic encoding \cite{joshi2021quantum, terhal2020towards}. However, their practical applications face a constraint in the form of the energy decay time, which is quantified by
the intrinsic quality factor $Q_\mathrm{i}$ of the resonator. A higher $Q_\mathrm{i}$ indicates longer coherence times, making its improvement a primary focus in the development of superconducting resonators for quantum technologies.

Three-dimensional cavity resonators, where the energy is stored within the dielectric medium inside the cavity (typically vacuum or a low-loss dielectric), have achieved exceptional $Q_\mathrm{i}$ as high as $10^8$ \cite{kudra2020high,heidler2021non,milul2023superconducting}, enabling breakthroughs such as break-even bosonic encoding for quantum error correction \cite{sivak2023real,ni2023beating}. Despite their performance, the large size and complexity of cavity resonators pose challenges for scalability.

In contrast, planar resonators---primarily two-dimensional conductor structures, such as microstrip lines or coplanar waveguides (CPWs)---are more scalable. They are fabricated on a substrate, with their electromagnetic fields concentrated between the resonator conductors, that is, in the gaps of a stripline or coplanar waveguide. The planar structure allows for fabrication using standard semiconductor processes such as photolithography and etching, and these resonators integrate seamlessly with planar superconducting qubits, making coupling straightforward. However, electromagnetic waves leaking into the substrate or surrounding space lead to radiation and dielectric losses, usually resulting in a lower $Q_\mathrm{i}$ than cavity resonators.
Recent advancements in surface treatment techniques \cite{goetz2016loss, vissers2010low, quintana2014characterization} and material investigation \cite{sage2011study,megrant2012planar, richardson2016fabrication} have shown promising results in enhancing the performance of planar resonators, with recent achievements reaching several million in the single-photon regime using epitaxial tantalum thin films \cite{shi2022tantalum}.

Hybrid designs aim to combine the high $Q_\mathrm{i}$ of 3D designs with the scalability of 2D structures \cite{lei2020high, ganjam2024surpassing}, where these resonators have essentially planar structures with additional 3D characteristics. For example, planar conductor patterns on a substrate combined with shielding or encapsulating structures to form a quasi-3D electromagnetic field distribution. While these hybrid architectures offer better performance than purely planar designs and maintain some degree of design flexibility, they still fall short in terms of the integration required for large-scale quantum computing systems. As such, the development of high-quality planar resonators remains critical for achieving fault-tolerant quantum computers. Optimization of the electric field distribution has been identified as a key factor in improving the $Q_\mathrm{i}$ of resonators.

A major limitation of planar resonators is energy dissipation due to two-level systems (TLSs) at material interfaces, which dominate losses at low temperatures and low power levels. The TLSs, which originate from material defects or disorder in the amorphous interface layer \cite{shalibo2010lifetime, martinis2005decoherence} forming tunneling states \cite{phillips1972tunneling}, can absorb energy from the resonator through capacitive coupling \cite{vissers2012identifying} and dissipate it to their own environments. TLS-induced losses not only limit qubit lifetimes and gate fidelities but also pose challenges for implementing bosonic codes, which require high-coherence resonators to encode quantum information in continuous-variable states. Consequently, understanding and mitigating TLS effects is crucial for improving planar resonator performance.

In this study, we explore alternative geometries to reduce coupling to TLSs in planar resonators, focusing on the use of Archimedean Spiral Resonators (ASRs). The spiral geometry distributes the electric field more evenly across the structure, reducing field concentration at the lossy interface and thus minimizing TLS-induced dissipation.
In addition, compared to CPWs, the high impedance of ASRs facilitates strong capacitive coupling to the qubit.
Unlike many complex resonator geometries, ASRs have well-established analytical models describing their resonance frequencies, impedance, and current distributions \cite{maleeva2015electrodynamics,peruzzo2020surpassing}, which allows for precise theoretical predictions and facilitates efficient design optimizations compared to structures requiring full numerical simulations.

To validate the advantage in $Q_\mathrm{i}$, we fabricated ASRs using domain-matched epitaxially grown titanium nitride (TiN) films on silicon wafers. Our measurements demonstrated intrinsic quality factors of $Q_\mathrm{i} = (9.6 \pm 1.5) \times 10^6$ at the single-photon level and $Q_\mathrm{i} = (9.91 \pm 0.39) \times 10^7$ at high power, significantly surpassing traditional CPW resonators.

We also conducted a comprehensive numerical analysis of the surface participation ratios (PRs) for both spiral resonators and CPWs using the finite element solver COMSOL \cite{comsol}. This analysis examined PR contributions from three critical interfaces: metal-air (MA), metal-substrate (MS), and substrate-air (SA), following methodologies widely employed in the study of microwave losses in superconducting materials \cite{wang2015surface, calusine2018analysis, woods2019determining, melville2020comparison, kudra2020high}. These calculations provide insight into the electromagnetic field distribution and energy storage mechanisms in these different resonator geometries.

The outline of this paper is as follows. In Section 2, we describe the design principles and theoretical foundations of coplanar waveguides and spiral resonators, including their frequency and impedance calculations, and detail our fabrication process and experimental setup. Section 3 presents our measurement results and analysis of the quality factors of the resonators. Finally, Section 4 examines the participation ratios to explain the superior performance of spiral geometries.

\section{Design and fabrication}
CPWs are the widely adopted choice for the design of superconducting planar resonators due to their simplicity and well-understood behavior. The characteristic frequency $f_\mathrm{CPW}$ and impedance $Z_\mathrm{CPW}$ of a CPW can be calculated using the parameters $k_1 = w/g$ and $k_2 = \tanh(\pi w/2h)/\tanh(\pi g/2h) $, where $w$ is the center conductor width, $g$ is the sum of the center conductor width plus the gaps on either side, and $h$ is the distance between the upper shielding and CPW. The equations for quarter-wavelength CPWs are as follows \cite{ghione1987coplanar}:
\begin{align}
  f_\mathrm{CPW} &= \frac{c}{4L\sqrt{\epsilon_\mathrm{eff}}} \label{eq1}\\
  Z_{\mathrm{CPW}} &= \frac{60\pi}{\sqrt{\epsilon_\mathrm{eff}}}
  \lr{\frac{K(k_1)}{K(k_1')} + \frac{K(k_2)}{K(k_2')}}^{-1}, \label{eq2}
\end{align}
where $c$ is the speed of light, $L$ is the length of the resonator, and $\epsilon_\mathrm{eff}$ is the effective dielectric constant calculated by
\begin{align}
  \epsilon_\mathrm{eff}= \lr{1 + \epsilon_\mathrm{sub}\frac{K(k_1')}{K(k_1)}\frac{K(k_2)}{K(k_2')}}
  \lr{1 + \frac{K(k_1')}{K(k_1)}\frac{K(k_2)}{K(k_2')}}^{-1}.
\end{align}
$K(k)$ is the complete elliptic integral of the first kind and $k_1' = \sqrt{1-k_1^2}$ and $ k_2' = \sqrt{1-k_2^2}$. These equations allow for precise design and optimization of CPW resonators for specific applications in quantum circuits.

The frequency of ASRs can be calculated by \cite{maleeva2015electrodynamics,peruzzo2020surpassing}
\begin{align}
  f_\mathrm{ASR} = \xi \frac{c}{\sqrt{\epsilon_\mathrm{eff}} } \frac{p}{ 2\pi (r_\mathrm{in} + np)^2}, \label{eq3}
\end{align}
where $\xi$ is a shape-dependent constant which is 0.81 for circular coils \cite{maleeva2015electrodynamics}, $r_\mathrm{in}$ is the inner radius, $n$ is the number of turns and $p$ is the pitch which is the wire width $w$ plus spacing. The effective dielectric constant $\epsilon_\mathrm{eff}$ in this case is apploximated as the average of the dielectric constants of air and substrate, $(1+\epsilon_\mathrm{sub})/2$.
The geometric inductance of ARSs can be calculated by the current-sheet method \cite{peruzzo2020surpassing,medahinne2024magnetic,mohan1999simple}
\begin{align}
  L = \frac{\mu_0  n^2  (r_\mathrm{out}+r_\mathrm{in})  c_1}{2}  \lr{\log(c_2/\rho) + c_3\rho + c_4\rho^2 },
\end{align}
where $\mu_0$ is the vacuum permeability, $r_\mathrm{out}$ is the outer radius and $\rho$ is the fill-ratio of the spiral given by $\rho = (r_\mathrm{out} - r_\mathrm{in})/(r_\mathrm{out} + r_\mathrm{in})$. $c_{1,2,3,4}$ are geometry dependent constants, $(1.0,2.5,0.0,0.2)$ for our case \cite{mohan1999simple}. Thus, the impedance can be calculated by
\begin{align}
  Z_\mathrm{ASR} = 2\pi f_\mathrm{ASR} L. \label{eq4}
\end{align}

\begin{figure}
  \centering
  \includegraphics[width=0.88\linewidth]{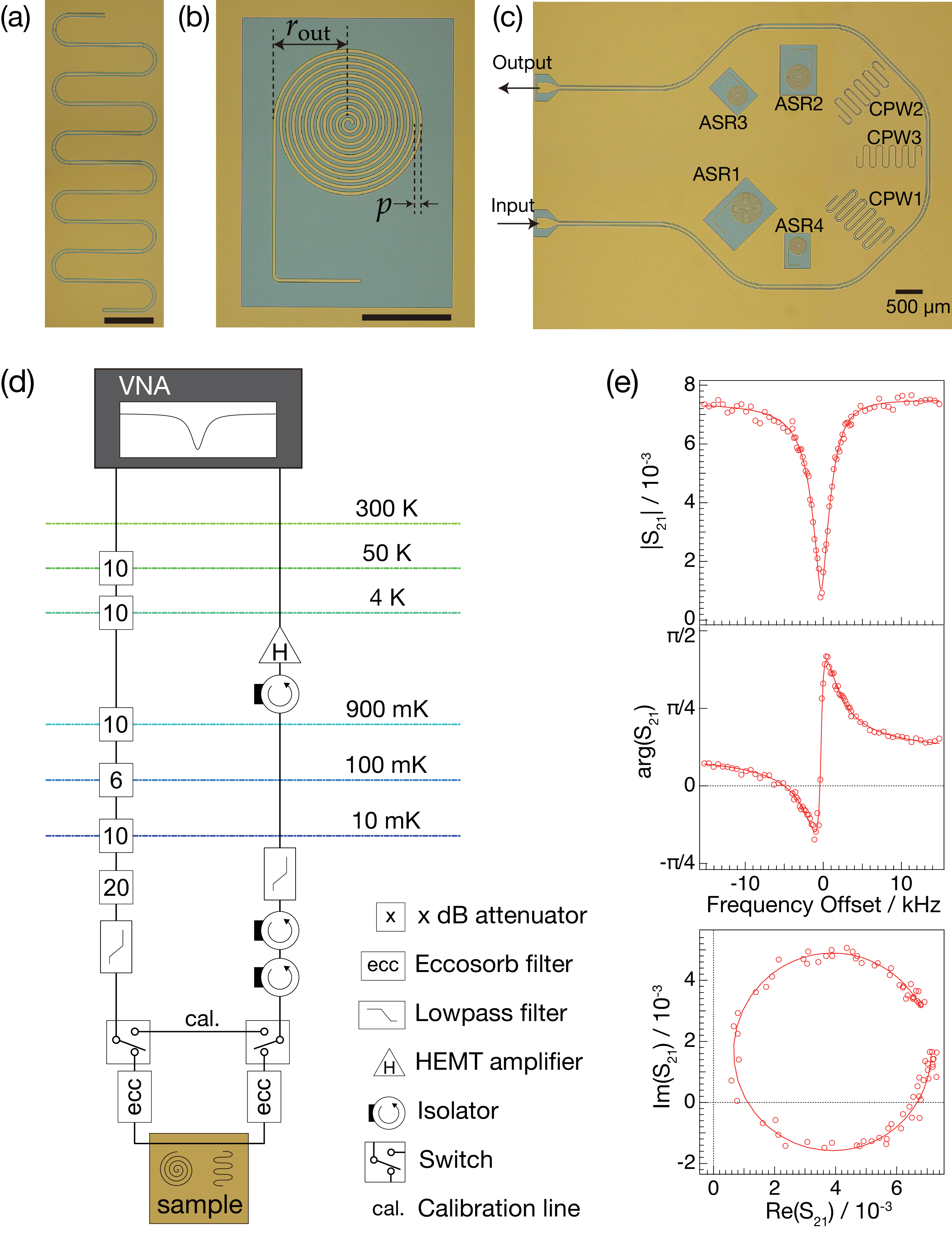}
  \caption{Experimental setup of the device. An optical microscope image of (a) a coplanar waveguide (CPW) quarter-wavelength resonator, and (b)  an Archimedean Spiral Resonator (ASR) with the outer radius $r_\mathrm{out}$ and the pitch $p$. The scale bars in (a) and (b) indicate 200 \textmu m, corresponding to CPW3 and ASR4 shown in (c).  (c) An optical image of the chip on which the resonators are coupled to the feedline. (d) Schematic of the measurement setup. The sample is bonded to a printed circuit board in a copper sample box and cooled to 10 mK in a dilution refrigerator. The incoming signal from the vector network analyzer (VNA) is attenuated by 75.8--79.0 dB (varying with frequency) to account for attenuators, cable losses, and bonding reflection. It then passes through an 8 GHz lowpass filter and an eccosorb filter. The output signal passes through another eccosorb filter, isolators, and a lowpass filter before low-temperature amplification with a high-electron-mobility transistor (HEMT).  Switches allow the signal to be routed either through the sample or directly through a calibration line for reference measurements. (e) A typical measured S$_{21}$ parameter of the resonator is shown here for ASR3 with an average photon number of 17.6 inside the resonator. The magnitude (top) and phase (middle) are mapped to the resonance circle (bottom), with the diameter $Q_l/|Q_c|$.}
  \label{fig0}
\end{figure}

In this work, we designed CPWs and ASRs with various line widths. The fundamental resonance frequencies were designed to range from 4 GHz to 7 GHz, based on Eqs.\ (\ref{eq1}) and (\ref{eq3}). The CPWs have line widths of $w =$ 8 \textmu m, 12 \textmu m, and 20 \textmu m, all with a quarter wavelength and a characteristic impedance of $50\ \Omega$ (Fig.\ \ref{fig0}(a) shows a CPW with $w = 8$ \textmu m). The ASRs share similar design characteristics but vary in their line widths and gaps. They have line widths of $w =$ 7 \textmu m, 8 \textmu m, 10 \textmu m, and 12 \textmu m, with gaps between neighboring wires matching these widths (i.e. $p = 2w$).
We fixed the number of spiral turns at 12, resulting in an outer radius of $r_\mathrm{out} = 12 p$ and a consistent impedance of $810\ \Omega$ for all ASRs, shown in Fig.\ \ref{fig0}(b).

These resonators, including both CPWs and ASRs, are fabricated on a single 10 mm $\times$ 10 mm chip and coupled to a common feedline as shown in Fig.\ \ref{fig0}(c). We implemented a design modification for the ASRs by extending the tail of the spiral shape to achieve capacitive coupling with the feedline. This extension decreases the fundamental frequencies of the ASRs compared to their pure spiral design, necessitating independent verification of the actual frequencies. The measured fundamental resonance frequencies of resonators with various designed width are presented in Table \ref{tab1}.

\begin{table}[]
  \caption{Resonator specifications—designed width ($w$), measured resonance frequency ($f$), and coupling quality factor ($Q_\mathrm{c}$).}
\begin{tabular*}{0.65\linewidth}{p{12ex}|p{15ex}p{15ex}p{15ex}} \hline\hline \label{tab1}
Label & $w$/ \textmu m & $f$/GHz & $Q_\mathrm{c} / 10^6$  \\ \hline\hline
CPW1  & 20             & 4.49    &    1.2                 \\ \hline
CPW2  & 12             & 5.34    &    1.4                 \\ \hline
CPW3  & 8              & 6.17    &    1.7                 \\ \hline \hline
ASR1  & 12             & 4.02    &    1.8                 \\ \hline
ASR2  & 10             & 4.81    &    1.0                 \\ \hline
ASR3  & 8              & 6.00    &    1.7                 \\ \hline
ASR4  & 7              & 6.85    &    1.6                 \\ \hline \hline
\end{tabular*}
\end{table}

The devices are fabricated on a (200)-oriented 100 nm TiN film domain-matched epitaxially grown on a high-resistive (20 $\mathrm{k\Omega\cdot cm}$) silicon (100) substrate \cite{sun2014fabrication}.
Fabrication was carried out via using photolithography followed by a hydrofluoric acid treatment to remove surface oxidation. The photolithography step employed a etching mask made with photoresist AZ1500 to pattern the TiN film through dry etching in CF$_4$ plasma. After removing the photoresist, O$_2$ ashing was conducted, and a time-controlled hydrofluoric acid treatment removed the native oxide layer. 
This process ensures clean interfaces and minimizes potential loss sources in the final resonator structures.

\section{Measurements and characterization}
We conducted measurements of the fabricated resonators using a dilution refrigerator at a base temperature of 10 mK. The sample was bonded to a printed circuit board in a copper sample box with an aluminum lid, creating a space of 12.3 mm $\times$ 12.3 mm $\times$ 1.5 mm, with a back-drill of radius 4.3 mm and depth of 3.1 mm. Our design minimized seams to reduce radiation loss, while the compact space and back-drill effectively raised the cut-off frequency of environmental modes.

Figure \ref{fig0}(d) illustrates the experimental setup. We utilized a series of attenuators 66 dB in total to reduce the incoming signal power. Taking into account cable losses and bonding reflections, the overall attenuation ranged from 75.8 to 79.0 dB, depending on the frequency (for a detailed breakdown, see Section \ref{sec1} of Supplemental Materials). The signal then passed through an 8 GHz lowpass filter and an eccosorb filter before reaching the sample. The resonators were probed using a vector network analyzer (VNA) to measure their transmission characteristics (S$_{21}$ parameters) over a range of input powers with a consistent resolution bandwidth of 10 Hz. The VNA output power varied from -80 dBm to 14 dBm, with an additional 16 dB attenuator placed at room temperature. Quality factors were extracted from the measured resonance curves using a circle fitting procedure \cite{probst2015efficient} that accounts for both intrinsic quality factor $Q_\mathrm{i}$ and coupling quality factor $Q_\mathrm{c}$.
Figure \ref{fig0}(e) shows a typical measured S$_{21}$ parameter of the ASR with $w=8$ \textmu m, mapped to the resonance circle with the diameter $Q_\mathrm{l}/|Q_\mathrm{c}|$, where $Q_\mathrm{l}$ is the load quality factor.
We designed the $Q_\mathrm{c}$ values to be approximately equal across resonators, and the actual results of $Q_\mathrm{c}$ from our measurements, shown in Table \ref{tab1}.

\begin{figure}
  \centering
  \includegraphics[width=\linewidth]{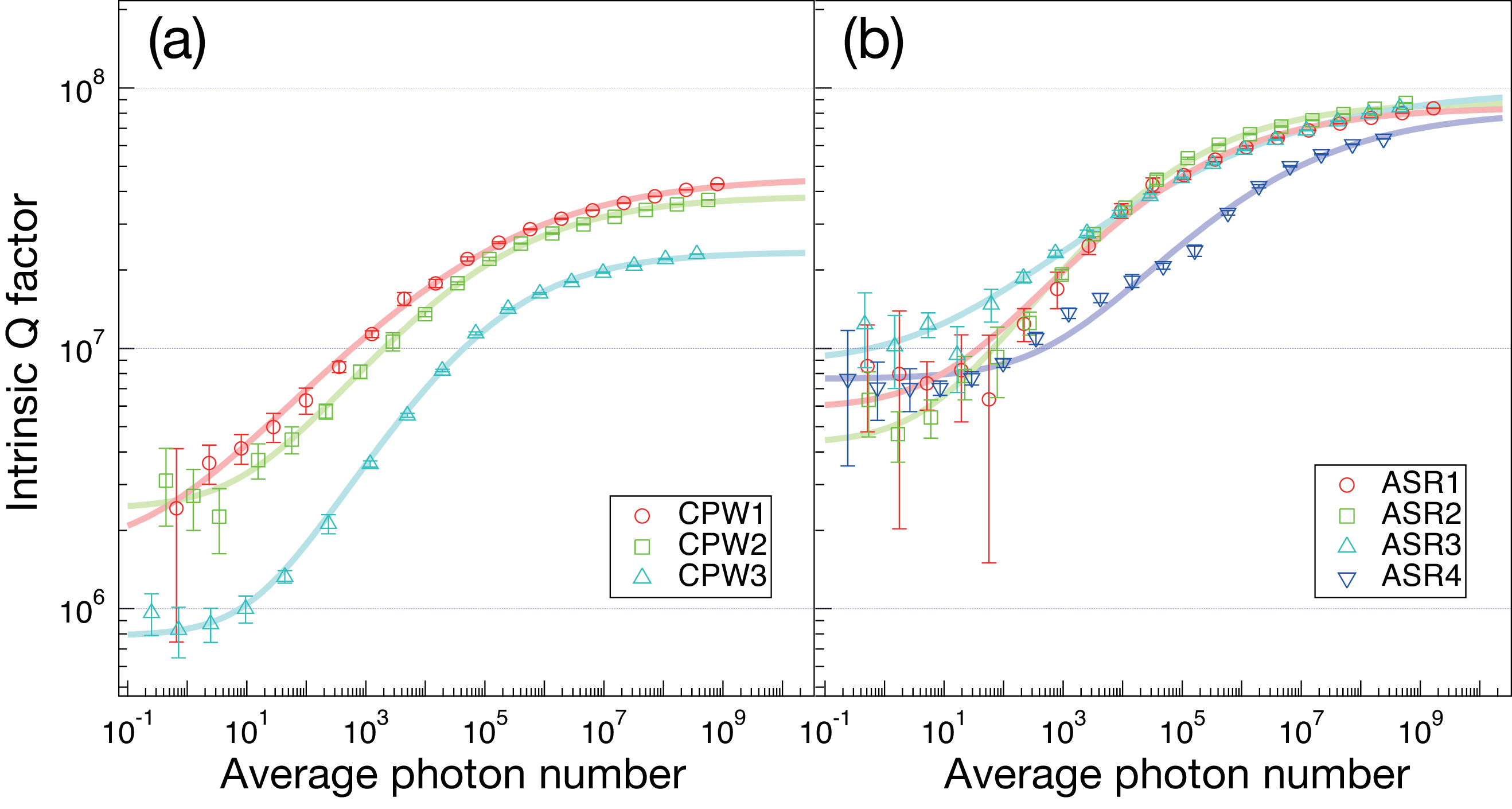}
  \caption{Extracted intrinsic quality factors of (a) CPWs and (b) ASRs. Markers are the measured quality factors, and lines are TLS fits using Eq.\ (\ref{eq10}).}
  \label{fig1}
\end{figure}
Extracted $Q_\mathrm{i}$ values for ASRs and CPWs are plotted in Fig.\ \ref{fig1}. Each data point represents the harmonic mean of ten individual measurements. Here, we observed a clear power dependence in the quality factors $Q_\mathrm{i}$, with the highest $Q_\mathrm{i}$ achieved at high input powers where the effects of two-level systems are saturated.
The power dependence of the $Q_\mathrm{i}$ of the resonator is described by the TLS model \cite{phillips1987two, wang2009improving}:
\begin{align}
\frac{1}{Q_\mathrm{i}} = \tan\delta \simeq p\, \delta_\mathrm{TLS} \frac{\tanh \lr{\frac{h f}{2 kT}}}{\sqrt{1+ \lr{\frac{\alr{n}}{n_\mathrm{c}}}^{\beta} } } + \delta_\mathrm{e}, \label{eq10}
\end{align}
where $p$ is the total energy participation ratio, $\delta_\mathrm{TLS}$ is the ensemble TLS loss tangent due to resonant absorption, $\alr{n}$ is the average photon number in resonators, $n_\mathrm{c}$ is the critical photon number at which TLS saturation occurs, and $\delta_\mathrm{e}$ is the power-independent contribution to the total loss. $\beta$ is an empirical parameter which describes TLS interaction. The measurement results are fit to Eq.\ (\ref{eq10}), allowing us to extract $p\, \delta_\mathrm{TLS}$ as shown in Table \ref{tab2}.

Our measurements revealed that the spiral resonators consistently outperformed their CPW counterparts across all power levels. At single-photon energies, the ASRs maintained intrinsic quality factors 2--4 times higher than CPWs of comparable dimensions. Our best-performing ASR exhibited an exceptional intrinsic quality factor reaching 10 million, with $Q_\mathrm{i,0} \equiv (p\, \delta_\mathrm{TLS})^{-1} = (9.6 \pm 1.5) \times 10^6$ at the single-photon level, and $Q_\mathrm{i,high} \equiv (\delta_\mathrm{e})^{-1} = (9.91 \pm 0.39) \times 10^7$ at high power. Table \ref{tab2} presents the single-photon intrinsic quality factor $Q_\mathrm{i,0}$ and high-power intrinsic quality factor $Q_\mathrm{i,high}$ for all resonators.


\begin{table}[]
  \caption{Measured and calculated parameters characterizing resonator performance. The loss tangent multiplied by surface participation ($p\, \delta_\mathrm{TLS}$), single-photon quality factor ($Q_\mathrm{i,0}$), and high-power quality factor ($Q_\mathrm{i,high}$) are obtained through transmission (S$_{21}$) measurements and TLS fits, with results shown in Fig.\ \ref{fig1}.
  The participation ratios of metal-air (MA), metal-substrate (MS), and substrate-air (SA) interfaces—$p_\mathrm{MA}$, $p_\mathrm{MS}$, and $p_\mathrm{SA}$, respectively—are calculated using the finite element solver COMSOL. The total surface participation ($p_\mathrm{tot}$) is the sum of $p_\mathrm{MA}$, $p_\mathrm{MS}$, and $p_\mathrm{SA}$.}
\begin{tabular*}{\linewidth}{p{7ex}|p{11ex}p{11ex}p{14ex}|p{9ex}p{9ex}p{9ex}p{10ex}} \hline\hline \label{tab2}
Label & $p\, \delta_\mathrm{TLS} / 10^{-8}$  & $Q_\mathrm{i,0} / 10^5$ & $Q_\mathrm{i,high} / 10^6$ & $ p_\mathrm{MA} / 10^{-5}$ & $ p_\mathrm{MS} / 10^{-5}$ & $ p_\mathrm{SA}/10^{-5}$ &$ p_\mathrm{tot}/10^{-5}$  \\ \hline\hline
CPW1  & 58  $\pm$ 31                                &  17.3  $\pm$ 9.3                 &   45.50 $\pm$ 0.88                    &    5.60                    &    56.50             &    54.81               &  116.91                 \\ \hline
CPW2  & 38.8 $\pm$ 5.7                                  &  25.8  $\pm$ 3.8                & 38.59 $\pm$ 0.59                   &    9.41                    &    88.42             &    87.90               &   185.73                \\ \hline
CPW3  & 123  $\pm$ 28                                &  8.13  $\pm$ 1.9                 &   23.46 $\pm$ 0.32                    &    14.50                    &    128.03             &   129.52                &  272.05                 \\ \hline \hline
ASR1  & 15.8  $\pm$ 4.5                                &  63 $\pm$ 18                   &   84.5  $\pm$ 2.3                   &    2.66              &   30.84              &    25.19               &      58.69             \\ \hline
ASR2  & 22.1 $\pm$ 6.2                                 &  45 $\pm$ 13                   &   88.0  $\pm$ 1.5                   &    3.06              &   34.31              &   28.82                &  66.19                 \\ \hline
ASR3  & 10.4  $\pm$ 1.6                               &  96 $\pm$15                   &   99.1  $\pm$ 3.9                  &    3.67             &  39.68               &   34.40                &  77.75                 \\ \hline
ASR4  & 11.9  $\pm$ 1.3                                &  84.0 $\pm$ 9.3                  &   81.8 $\pm$ 6.3                   &    4.12              &    43.59             &   38.44                &    86.15               \\ \hline \hline
\end{tabular*}
\end{table}

\section{Surface participation ratio}
To elucidate the superior performance of ASRs, we computed the surface participation ratios (PRs) of resonators. We categorized the key interfaces into three types: metal-air (MA), metal-substrate (MS), and substrate-air (SA). By analyzing these interfaces separately, we assessed their individual contributions to loss. The total loss tangent associated with these interfaces is given by
\begin{align}
  \tan \delta = \sum_i p_i \tan \delta_i,
\end{align}
where interface type $i = \text{MA, MS, or SA}$ has loss tangent $\delta_i$ and PR quantifies the fraction of the electromagnetic energy $U_i$ stored at amorphous interfaces relative to the total energy $U_\mathrm{tot}$:
\begin{align}
  p_i = \frac{U_i}{U_\mathrm{tot}} = \int_i dxdydz\ \frac{\epsilon_i}{2} \abs{\E(x,y,z)}^2 / U_\mathrm{tot} \label{eq9}
\end{align}
for the interface material with dielectric constant $\epsilon_i$.

\begin{figure}
  \centering
  \includegraphics[width=\linewidth]{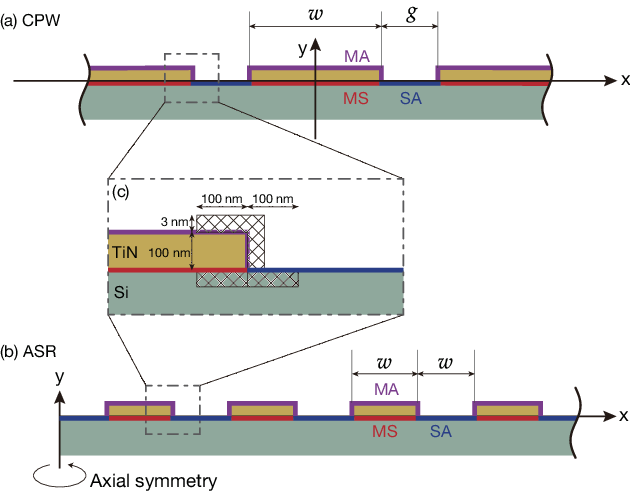}
  \caption{The schematic illustration of the model for 2D simulation of PRs shows: (a) a CPW with width $w$ and gap $g$, and (b) an ASR with width $w$ and spacing $w$. The ASR is approximated as a set of concentric rings using axially symmetric modeling. The three interfaces are depicted in purple (MA), red (MS), and blue (SA). (c) A detailed model of the metal edge, where a 3 nm layer within 100 nm of the peripheral region is modeled, as indicated by the cross-hatched area.}
  \label{fig2}
\end{figure}

We conducted 2D static calculations using COMSOL. For CPWs, we adopted a calculation method similar to that described in Ref.\ \cite{wenner2011surface}, modeling them as infinitely long rectangular metal plates, as shown in Fig.\ \ref{fig2}(a). ASRs were approximated as sets of concentric rings using axially symmetric modeling, as shown in Fig.\ \ref{fig2}(b).

In addition, we introduce an approximation that the electric field $\E$ is uniform over the thin lossy layer, allowing us to calculate the PR solely from the electric field on the surface. This approximation holds for most of the structure, except for the peripheral region of the metal where field variations are more significant \cite{wang2015surface, wenner2011surface}.
When we take the $x$-axis parallel to the surface of the substrate and the $y$-axis perpendicular to it, the continuity of electric displacement requires:
\begin{align}
  E_x^{\mathrm{sim},i} &= E_x^i\\
  \epsilon_{\mathrm{sim},i} E_y^{\mathrm{sim},i} &= \epsilon_i E_y^i,
\end{align}
where the subscript $i = $ MA, MS, or SA refers to the lossy interface the actual sample contains and `sim,$i$' the material present in the simulation instead of the actual material of $i$. Here, we supposed $\epsilon_i = 10$ for all $i$ and we used $\epsilon_{\mathrm{sim,MA}} = \epsilon_\mathrm{air} = 1$, $\epsilon_{\mathrm{sim,MS}} = \epsilon_\mathrm{Si} = 11.45$
and $\epsilon_{\mathrm{sim,SA}} = (\epsilon_\mathrm{air}+\epsilon_\mathrm{Si})/2 = 6.225$ for the relative dielectric constant.
Thus, the PR in the internal region of the metal should be calculated by
\begin{align}
  p_i = t \int_i dx dy\ \frac{\epsilon_i}{2} \lr{ \abs{E_x^{\mathrm{sim},i}}^2 + \abs{\frac{\epsilon_{\mathrm{sim},i}}{\epsilon_i}E_y^{\mathrm{sim},i}(x)}^2 }/U_\mathrm{tot},
\end{align}
where $t$ represents the thickness of the hypothetical lossy layer, here we supposed $t = 3$ nm.

Using the continuity conditions of electric displacement, we can calculate the electric field in the thin layer without explicitly modeling it in internal regions, significantly reducing computational costs. However, the peripheral region still requires detailed modeling due to its more complex field distribution. We modeled the 3 nm layer within 100 nm of the peripheral region as shown in Fig.\ \ref{fig2}(c) and calculated PRs using Eq.\ (\ref{eq9}).

The voltage distribution $V(x)$ in an ASR can be expressed as a function of the radial distance $x$ from the center:
\begin{align}
  V(x) = V_0 \cos\lr{\pi \lr{\frac{x}{r_\mathrm{out}}}^2 } \label{eq13}
\end{align}
where $V_0$ is the applied voltage and $r_\mathrm{out}$ is the outer radius of the spiral. This analytical solution allows for efficient calculation of the electric field distribution, which is crucial for accurate PR calculations. The validity of Eq.\ (\ref{eq13}) was checked using COMSOL simulation (see Section \ref{sec3} of Supplemental Materials).

Our COMSOL simulations revealed that ASRs have significantly lower PRs compared to CPWs, as presented in Table \ref{tab2}. The total surface participation ($p_\mathrm{tot}$) is the sum of $p_\mathrm{MA}$, $p_\mathrm{MS}$, and $p_\mathrm{SA}$. This reduction in PR largely explains the superior $Q_\mathrm{i}$ values of ASRs, even when considering coarse approximations such as modeling a spiral with rings.

\begin{figure}
  \centering
  \includegraphics[width=0.7\linewidth]{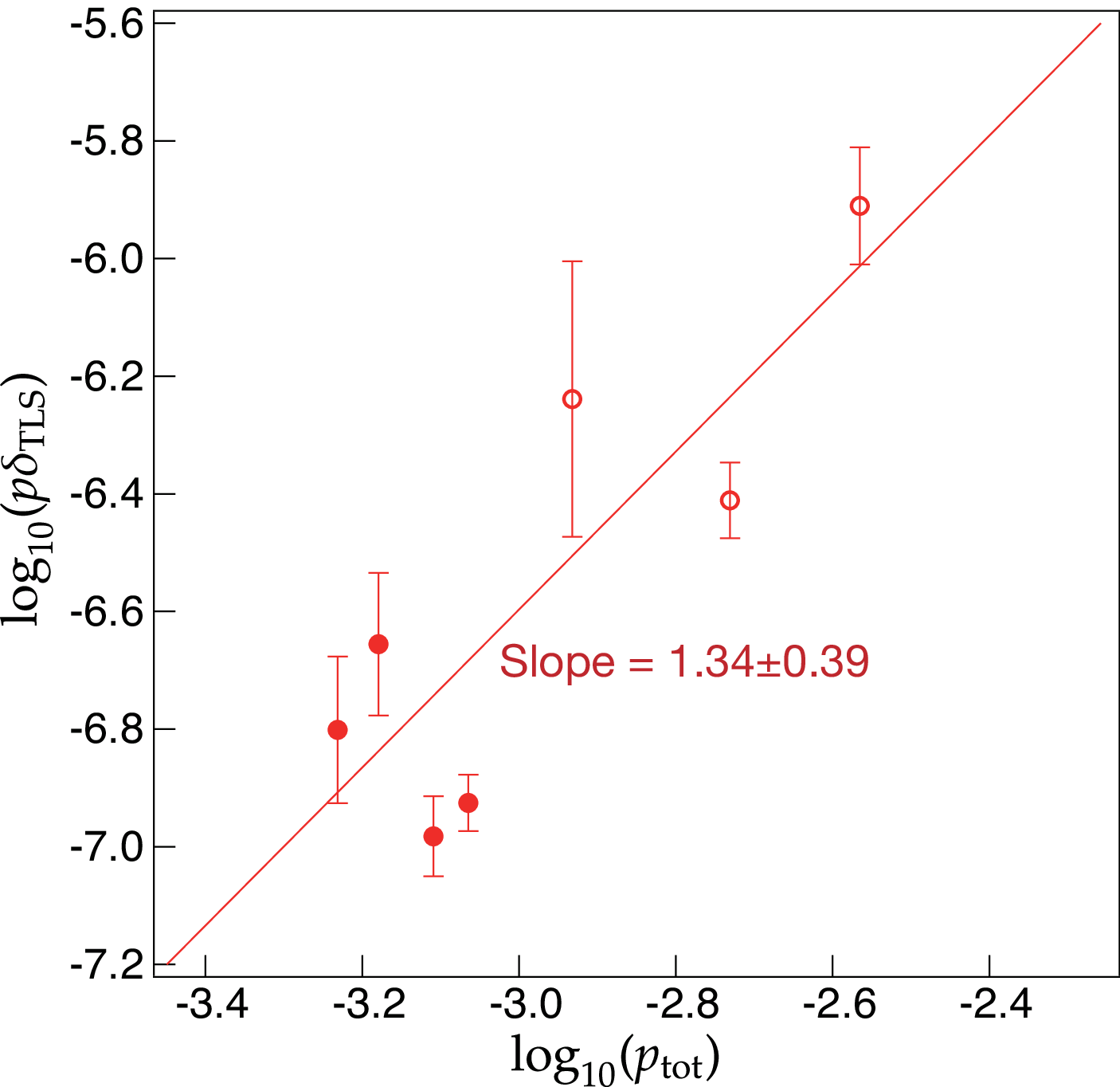}
  \caption{The relationship between experimentally extracted loss tangent values $p\, \delta_\mathrm{TLS}$ and calculated total participation ratios $p_\mathrm{tot}$. The vertical axis represents $\log(p \tan \delta)$ while the horizontal axis represents $\log(p_\mathrm{tot})$. Filled circles correspond to ASR data, while open circles represent CPW data. A linear fit to the data yields a slope of $1.34 \pm 0.39$. The error bars represent fitting errors obtained from Fig.\ \ref{fig1}, propagated through logarithmic transformation.
  }
  \label{fig4}
\end{figure}
Figure \ref{fig4} shows the relationship between the experimentally extracted and calculated loss tangent on a log-log scale. Because the error distribution of the loss tangent follows a log-normal distribution (see Section \ref{sec2} of Supplemental Materials), a log-log representation provides a more appropriate visualization. At low power levels, the loss tangent fluctuates due to variations in the TLS state, as shown in Fig.\ \ref{fig_s2}, occasionally leading to large experimental values. A linear fit to the data yields a slope of $1.34\pm 0.39$, which, given the uncertainty, is consistent with the ideal value of unity. This agreement confirms the validity of the PR simulation, as a slope near unity indicates that the model accurately captures the dominant loss mechanisms. The slight deviation from unity suggests that additional loss sources, such as bulk loss \cite{zhang2024acceptor} or other surface-related effects \cite{de2020two}, may also contribute, though they do not dominate the behavior.

 The decreased surface participation directly correlates with the observed reduction in TLS induced losses, resulting in enhanced resonator performance. The MS and SA interfaces still contribute significantly to loss. Future designs that reduce the PR at these interfaces could lead to substantial improvements in resonator performance.

\section{Conclusion}
In conclusion, our study demonstrates the potential of spiral geometry in superconducting resonators for achieving high intrinsic quality factors. Using domain-matched epitaxially grown titanium nitride on silicon wafers, we have fabricated Archimedean spiral resonators with high quality factors, achieving $Q_\mathrm{i} = (9.6 \pm 1.5) \times 10^6$ at the single-photon level and $Q_\mathrm{i} = (9.91 \pm 0.39) \times 10^7$ at high power. Our numerical analysis revealed that the superior performance of our spiral resonators compared to traditional coplanar waveguide resonators is explained by their lower participation ratios at lossy interfaces, which result from a more evenly distributed electromagnetic field. These results suggest that spiral resonators could be a promising alternative to traditional coplanar waveguide designs, particularly in applications requiring minimal coupling to two-level systems, such as quantum computing and quantum sensing. Furthermore, the high-$Q_\mathrm{i}$ performance of our spiral resonators may provide a platform for studying the dynamics of two-level systems.

While these results are promising, the metal-substrate and substrate-air interfaces still contribute significantly to losses. Future work could explore further optimizations of the spiral geometry, such as varying the width along the spiral or incorporating other design elements to reduce surface participation further. Additionally, investigating the performance of these spiral resonators in different material systems such as niobium, aluminum, or tantalum could provide deeper insights into the material-dependent losses \cite{melville2020comparison,wang2022towards}. Exploring alternative wafer materials, such as sapphire \cite{deng2023titanium} or silicon-on-insulator \cite{keller2017transmon}, could also provide insights into the role of substrate-induced losses and help identify the fundamental limits of superconducting resonators for further improvement.

These improvements address the challenge of enhancing the scalability and reliability of superconducting quantum circuits, including the implementation of bosonic code quantum error correction, and may also contribute to advances in quantum sensing, such as dark matter searches with qubits \cite{dixit2021searching}.

\backmatter





\section*{Declarations}
\bmhead{Acknowledgements}
The authors acknowledge the Superconducting Quantum Electronics Research Team, the Superconducting Quantum Computing System Research Unit, and the Semiconductor Science Research Support Team for their support in device fabrication at the RIKEN Nanoscience Joint Laboratory. We also thank G. Ando, Y. Matsuyama, and Y. Tsuchimoto for their assistance in experiments, and S. Tamate for valuable insights into the PR calculation methods.

\bmhead{Funding}
This research has been supported by funding from JST Moonshot R\&D Program (Grant Number JPMJMS2067), JST ERATO (Grant Number JPMJER2302), JST CREST (Grant Number JPMJCR24I5), MEXT Q-LEAP (Grant Number JPMXS0118068682), and JSPS KAKENHI (Grant Numbers JP24H00832 and JP24K22871).

\bmhead{Data availability}
The datasets used and/or analysed during the current study are available from the corresponding author on reasonable request.

\bmhead{Materials availability}
Not applicable.

\bmhead{Code availability}
Not applicable.

\bmhead{Ethics approval and consent to participate}
Not applicable.

\bmhead{Consent for publication}
Not applicable

\bmhead{Competing interests}
The authors declare no competing interests.

\bmhead{Author contribution}
This project was conceptualized and led by A.N., S.S., and Y.T. Y.H, and H.T prepared domain-matched epitaxially grown TiN films. Y.T. carried out sample design, fabrication, measurements, and data analysis, with support from S.S. Y.T. drafted the manuscript, which was reviewed and approved by all authors.

\bibliography{mybib}

\begin{thebibliography}{10}
\expandafter\ifx\csname url\endcsname\relax
  \def\url#1{\burl{#1}}\fi
\expandafter\ifx\csname urlprefix\endcsname\relax\def\urlprefix{URL }\fi
\providecommand{\bibinfo}[2]{#2}
\providecommand{\eprint}[2][]{\url{#2}}
\providecommand{\doi}[1]{\url{https://doi.org/#1}}
\bibcommenthead

\bibitem{blais2004cavity}
\bibinfo{author}{Blais, A.}, \bibinfo{author}{Huang, R.-S.},
  \bibinfo{author}{Wallraff, A.}, \bibinfo{author}{Girvin, S.~M.} \&
  \bibinfo{author}{Schoelkopf, R.~J.}
\newblock \bibinfo{title}{Cavity quantum electrodynamics for superconducting
  electrical circuits: An architecture for quantum computation}.
\newblock \emph{\bibinfo{journal}{Physical Review A—Atomic, Molecular, and
  Optical Physics}} \textbf{\bibinfo{volume}{69}}, \bibinfo{pages}{062320}
  (\bibinfo{year}{2004}).

\bibitem{wallraff2004strong}
\bibinfo{author}{Wallraff, A.} \emph{et~al.}
\newblock \bibinfo{title}{Strong coupling of a single photon to a
  superconducting qubit using circuit quantum electrodynamics}.
\newblock \emph{\bibinfo{journal}{Nature}} \textbf{\bibinfo{volume}{431}},
  \bibinfo{pages}{162--167} (\bibinfo{year}{2004}).

\bibitem{sillanpaa2007coherent}
\bibinfo{author}{Sillanp{\"a}{\"a}, M.~A.}, \bibinfo{author}{Park, J.~I.} \&
  \bibinfo{author}{Simmonds, R.~W.}
\newblock \bibinfo{title}{Coherent quantum state storage and transfer between
  two phase qubits via a resonant cavity}.
\newblock \emph{\bibinfo{journal}{Nature}} \textbf{\bibinfo{volume}{449}},
  \bibinfo{pages}{438--442} (\bibinfo{year}{2007}).

\bibitem{bergeal2010phase}
\bibinfo{author}{Bergeal, N.} \emph{et~al.}
\newblock \bibinfo{title}{Phase-preserving amplification near the quantum limit
  with a josephson ring modulator}.
\newblock \emph{\bibinfo{journal}{Nature}} \textbf{\bibinfo{volume}{465}},
  \bibinfo{pages}{64--68} (\bibinfo{year}{2010}).

\bibitem{day2003broadband}
\bibinfo{author}{Day, P.~K.}, \bibinfo{author}{LeDuc, H.~G.},
  \bibinfo{author}{Mazin, B.~A.}, \bibinfo{author}{Vayonakis, A.} \&
  \bibinfo{author}{Zmuidzinas, J.}
\newblock \bibinfo{title}{A broadband superconducting detector suitable for use
  in large arrays}.
\newblock \emph{\bibinfo{journal}{Nature}} \textbf{\bibinfo{volume}{425}},
  \bibinfo{pages}{817--821} (\bibinfo{year}{2003}).

\bibitem{hofheinz2009synthesizing}
\bibinfo{author}{Hofheinz, M.} \emph{et~al.}
\newblock \bibinfo{title}{Synthesizing arbitrary quantum states in a
  superconducting resonator}.
\newblock \emph{\bibinfo{journal}{Nature}} \textbf{\bibinfo{volume}{459}},
  \bibinfo{pages}{546--549} (\bibinfo{year}{2009}).

\bibitem{joshi2021quantum}
\bibinfo{author}{Joshi, A.}, \bibinfo{author}{Noh, K.} \& \bibinfo{author}{Gao,
  Y.~Y.}
\newblock \bibinfo{title}{Quantum information processing with bosonic qubits in
  circuit qed}.
\newblock \emph{\bibinfo{journal}{Quantum Science and Technology}}
  \textbf{\bibinfo{volume}{6}}, \bibinfo{pages}{033001} (\bibinfo{year}{2021}).

\bibitem{terhal2020towards}
\bibinfo{author}{Terhal, B.~M.}, \bibinfo{author}{Conrad, J.} \&
  \bibinfo{author}{Vuillot, C.}
\newblock \bibinfo{title}{Towards scalable bosonic quantum error correction}.
\newblock \emph{\bibinfo{journal}{Quantum Science and Technology}}
  \textbf{\bibinfo{volume}{5}}, \bibinfo{pages}{043001} (\bibinfo{year}{2020}).

\bibitem{kudra2020high}
\bibinfo{author}{Kudra, M.} \emph{et~al.}
\newblock \bibinfo{title}{High quality three-dimensional aluminum microwave
  cavities}.
\newblock \emph{\bibinfo{journal}{Applied Physics Letters}}
  \textbf{\bibinfo{volume}{117}} (\bibinfo{year}{2020}).

\bibitem{heidler2021non}
\bibinfo{author}{Heidler, P.} \emph{et~al.}
\newblock \bibinfo{title}{Non-markovian effects of two-level systems in a
  niobium coaxial resonator with a single-photon lifetime of 10 milliseconds}.
\newblock \emph{\bibinfo{journal}{Physical Review Applied}}
  \textbf{\bibinfo{volume}{16}}, \bibinfo{pages}{034024}
  (\bibinfo{year}{2021}).

\bibitem{milul2023superconducting}
\bibinfo{author}{Milul, O.} \emph{et~al.}
\newblock \bibinfo{title}{Superconducting cavity qubit with tens of
  milliseconds single-photon coherence time}.
\newblock \emph{\bibinfo{journal}{PRX Quantum}} \textbf{\bibinfo{volume}{4}},
  \bibinfo{pages}{030336} (\bibinfo{year}{2023}).

\bibitem{sivak2023real}
\bibinfo{author}{Sivak, V.} \emph{et~al.}
\newblock \bibinfo{title}{Real-time quantum error correction beyond
  break-even}.
\newblock \emph{\bibinfo{journal}{Nature}} \textbf{\bibinfo{volume}{616}},
  \bibinfo{pages}{50--55} (\bibinfo{year}{2023}).

\bibitem{ni2023beating}
\bibinfo{author}{Ni, Z.} \emph{et~al.}
\newblock \bibinfo{title}{Beating the break-even point with a
  discrete-variable-encoded logical qubit}.
\newblock \emph{\bibinfo{journal}{Nature}} \textbf{\bibinfo{volume}{616}},
  \bibinfo{pages}{56--60} (\bibinfo{year}{2023}).

\bibitem{goetz2016loss}
\bibinfo{author}{Goetz, J.} \emph{et~al.}
\newblock \bibinfo{title}{Loss mechanisms in superconducting thin film
  microwave resonators}.
\newblock \emph{\bibinfo{journal}{Journal of Applied Physics}}
  \textbf{\bibinfo{volume}{119}} (\bibinfo{year}{2016}).

\bibitem{vissers2010low}
\bibinfo{author}{Vissers, M.~R.} \emph{et~al.}
\newblock \bibinfo{title}{Low loss superconducting titanium nitride coplanar
  waveguide resonators}.
\newblock \emph{\bibinfo{journal}{Applied Physics Letters}}
  \textbf{\bibinfo{volume}{97}} (\bibinfo{year}{2010}).

\bibitem{quintana2014characterization}
\bibinfo{author}{Quintana, C.} \emph{et~al.}
\newblock \bibinfo{title}{Characterization and reduction of
  microfabrication-induced decoherence in superconducting quantum circuits}.
\newblock \emph{\bibinfo{journal}{Applied Physics Letters}}
  \textbf{\bibinfo{volume}{105}} (\bibinfo{year}{2014}).

\bibitem{sage2011study}
\bibinfo{author}{Sage, J.~M.}, \bibinfo{author}{Bolkhovsky, V.},
  \bibinfo{author}{Oliver, W.~D.}, \bibinfo{author}{Turek, B.} \&
  \bibinfo{author}{Welander, P.~B.}
\newblock \bibinfo{title}{Study of loss in superconducting coplanar waveguide
  resonators}.
\newblock \emph{\bibinfo{journal}{Journal of Applied Physics}}
  \textbf{\bibinfo{volume}{109}} (\bibinfo{year}{2011}).

\bibitem{megrant2012planar}
\bibinfo{author}{Megrant, A.} \emph{et~al.}
\newblock \bibinfo{title}{Planar superconducting resonators with internal
  quality factors above one million}.
\newblock \emph{\bibinfo{journal}{Applied Physics Letters}}
  \textbf{\bibinfo{volume}{100}} (\bibinfo{year}{2012}).

\bibitem{richardson2016fabrication}
\bibinfo{author}{Richardson, C.~J.} \emph{et~al.}
\newblock \bibinfo{title}{Fabrication artifacts and parallel loss channels in
  metamorphic epitaxial aluminum superconducting resonators}.
\newblock \emph{\bibinfo{journal}{Superconductor Science and Technology}}
  \textbf{\bibinfo{volume}{29}}, \bibinfo{pages}{064003}
  (\bibinfo{year}{2016}).

\bibitem{shi2022tantalum}
\bibinfo{author}{Shi, L.} \emph{et~al.}
\newblock \bibinfo{title}{Tantalum microwave resonators with ultra-high
  intrinsic quality factors}.
\newblock \emph{\bibinfo{journal}{Applied Physics Letters}}
  \textbf{\bibinfo{volume}{121}} (\bibinfo{year}{2022}).

\bibitem{lei2020high}
\bibinfo{author}{Lei, C.~U.}, \bibinfo{author}{Krayzman, L.},
  \bibinfo{author}{Ganjam, S.}, \bibinfo{author}{Frunzio, L.} \&
  \bibinfo{author}{Schoelkopf, R.~J.}
\newblock \bibinfo{title}{High coherence superconducting microwave cavities
  with indium bump bonding}.
\newblock \emph{\bibinfo{journal}{Applied Physics Letters}}
  \textbf{\bibinfo{volume}{116}} (\bibinfo{year}{2020}).

\bibitem{ganjam2024surpassing}
\bibinfo{author}{Ganjam, S.} \emph{et~al.}
\newblock \bibinfo{title}{Surpassing millisecond coherence in on chip
  superconducting quantum memories by optimizing materials and circuit design}.
\newblock \emph{\bibinfo{journal}{Nature Communications}}
  \textbf{\bibinfo{volume}{15}}, \bibinfo{pages}{3687} (\bibinfo{year}{2024}).

\bibitem{shalibo2010lifetime}
\bibinfo{author}{Shalibo, Y.} \emph{et~al.}
\newblock \bibinfo{title}{Lifetime and coherence of two-level defects in a
  josephson junction}.
\newblock \emph{\bibinfo{journal}{Physical review letters}}
  \textbf{\bibinfo{volume}{105}}, \bibinfo{pages}{177001}
  (\bibinfo{year}{2010}).

\bibitem{martinis2005decoherence}
\bibinfo{author}{Martinis, J.~M.} \emph{et~al.}
\newblock \bibinfo{title}{Decoherence in josephson qubits from dielectric
  loss}.
\newblock \emph{\bibinfo{journal}{Physical review letters}}
  \textbf{\bibinfo{volume}{95}}, \bibinfo{pages}{210503}
  (\bibinfo{year}{2005}).

\bibitem{phillips1972tunneling}
\bibinfo{author}{Phillips, W.~A.}
\newblock \bibinfo{title}{Tunneling states in amorphous solids}.
\newblock \emph{\bibinfo{journal}{Journal of low temperature physics}}
  \textbf{\bibinfo{volume}{7}}, \bibinfo{pages}{351--360}
  (\bibinfo{year}{1972}).

\bibitem{vissers2012identifying}
\bibinfo{author}{Vissers, M.~R.}, \bibinfo{author}{Weides, M.~P.},
  \bibinfo{author}{Kline, J.~S.}, \bibinfo{author}{Sandberg, M.} \&
  \bibinfo{author}{Pappas, D.~P.}
\newblock \bibinfo{title}{Identifying capacitive and inductive loss in lumped
  element superconducting hybrid titanium nitride/aluminum resonators}.
\newblock \emph{\bibinfo{journal}{Applied Physics Letters}}
  \textbf{\bibinfo{volume}{101}} (\bibinfo{year}{2012}).

\bibitem{maleeva2015electrodynamics}
\bibinfo{author}{Maleeva, N.} \emph{et~al.}
\newblock \bibinfo{title}{Electrodynamics of planar archimedean spiral
  resonator}.
\newblock \emph{\bibinfo{journal}{Journal of Applied Physics}}
  \textbf{\bibinfo{volume}{118}} (\bibinfo{year}{2015}).

\bibitem{peruzzo2020surpassing}
\bibinfo{author}{Peruzzo, M.}, \bibinfo{author}{Trioni, A.},
  \bibinfo{author}{Hassani, F.}, \bibinfo{author}{Zemlicka, M.} \&
  \bibinfo{author}{Fink, J.~M.}
\newblock \bibinfo{title}{Surpassing the resistance quantum with a geometric
  superinductor}.
\newblock \emph{\bibinfo{journal}{Physical Review Applied}}
  \textbf{\bibinfo{volume}{14}}, \bibinfo{pages}{044055}
  (\bibinfo{year}{2020}).

\bibitem{comsol}
\emph{\bibinfo{journal}{COMSOL 6.0}}  (\bibinfo{year}{www.comsol.com}).

\bibitem{wang2015surface}
\bibinfo{author}{Wang, C.} \emph{et~al.}
\newblock \bibinfo{title}{Surface participation and dielectric loss in
  superconducting qubits}.
\newblock \emph{\bibinfo{journal}{Applied Physics Letters}}
  \textbf{\bibinfo{volume}{107}} (\bibinfo{year}{2015}).

\bibitem{calusine2018analysis}
\bibinfo{author}{Calusine, G.} \emph{et~al.}
\newblock \bibinfo{title}{Analysis and mitigation of interface losses in
  trenched superconducting coplanar waveguide resonators}.
\newblock \emph{\bibinfo{journal}{Applied Physics Letters}}
  \textbf{\bibinfo{volume}{112}} (\bibinfo{year}{2018}).

\bibitem{woods2019determining}
\bibinfo{author}{Woods, W.} \emph{et~al.}
\newblock \bibinfo{title}{Determining interface dielectric losses in
  superconducting coplanar-waveguide resonators}.
\newblock \emph{\bibinfo{journal}{Physical Review Applied}}
  \textbf{\bibinfo{volume}{12}}, \bibinfo{pages}{014012}
  (\bibinfo{year}{2019}).

\bibitem{melville2020comparison}
\bibinfo{author}{Melville, A.} \emph{et~al.}
\newblock \bibinfo{title}{Comparison of dielectric loss in titanium nitride and
  aluminum superconducting resonators}.
\newblock \emph{\bibinfo{journal}{Applied Physics Letters}}
  \textbf{\bibinfo{volume}{117}} (\bibinfo{year}{2020}).

\bibitem{ghione1987coplanar}
\bibinfo{author}{Ghione, G.} \& \bibinfo{author}{Naldi, C.~U.}
\newblock \bibinfo{title}{Coplanar waveguides for mmic applications: Effect of
  upper shielding, conductor backing, finite-extent ground planes, and
  line-to-line coupling}.
\newblock \emph{\bibinfo{journal}{IEEE transactions on Microwave Theory and
  Techniques}} \textbf{\bibinfo{volume}{35}}, \bibinfo{pages}{260--267}
  (\bibinfo{year}{1987}).

\bibitem{medahinne2024magnetic}
\bibinfo{author}{Medahinne, M.} \emph{et~al.}
\newblock \bibinfo{title}{Magnetic field tolerant superconducting spiral
  resonators for circuit qed}.
\newblock \emph{\bibinfo{journal}{arXiv preprint arXiv:2406.10386}}
  (\bibinfo{year}{2024}).

\bibitem{mohan1999simple}
\bibinfo{author}{Mohan, S.~S.}, \bibinfo{author}{del Mar~Hershenson, M.},
  \bibinfo{author}{Boyd, S.~P.} \& \bibinfo{author}{Lee, T.~H.}
\newblock \bibinfo{title}{Simple accurate expressions for planar spiral
  inductances}.
\newblock \emph{\bibinfo{journal}{IEEE Journal of solid-state circuits}}
  \textbf{\bibinfo{volume}{34}}, \bibinfo{pages}{1419--1424}
  (\bibinfo{year}{1999}).

\bibitem{sun2014fabrication}
\bibinfo{author}{Sun, R.}, \bibinfo{author}{Makise, K.}, \bibinfo{author}{Qiu,
  W.}, \bibinfo{author}{Terai, H.} \& \bibinfo{author}{Wang, Z.}
\newblock \bibinfo{title}{Fabrication of (200)-oriented tin films on si (100)
  substrates by dc magnetron sputtering}.
\newblock \emph{\bibinfo{journal}{IEEE Transactions on Applied
  Superconductivity}} \textbf{\bibinfo{volume}{25}}, \bibinfo{pages}{1--4}
  (\bibinfo{year}{2014}).

\bibitem{probst2015efficient}
\bibinfo{author}{Probst, S.}, \bibinfo{author}{Song, F.},
  \bibinfo{author}{Bushev, P.~A.}, \bibinfo{author}{Ustinov, A.~V.} \&
  \bibinfo{author}{Weides, M.}
\newblock \bibinfo{title}{Efficient and robust analysis of complex scattering
  data under noise in microwave resonators}.
\newblock \emph{\bibinfo{journal}{Review of Scientific Instruments}}
  \textbf{\bibinfo{volume}{86}} (\bibinfo{year}{2015}).

\bibitem{phillips1987two}
\bibinfo{author}{Phillips, W.~A.}
\newblock \bibinfo{title}{Two-level states in glasses}.
\newblock \emph{\bibinfo{journal}{Reports on Progress in Physics}}
  \textbf{\bibinfo{volume}{50}}, \bibinfo{pages}{1657} (\bibinfo{year}{1987}).

\bibitem{wang2009improving}
\bibinfo{author}{Wang, H.} \emph{et~al.}
\newblock \bibinfo{title}{Improving the coherence time of superconducting
  coplanar resonators}.
\newblock \emph{\bibinfo{journal}{Applied Physics Letters}}
  \textbf{\bibinfo{volume}{95}} (\bibinfo{year}{2009}).

\bibitem{wenner2011surface}
\bibinfo{author}{Wenner, J.} \emph{et~al.}
\newblock \bibinfo{title}{Surface loss simulations of superconducting coplanar
  waveguide resonators}.
\newblock \emph{\bibinfo{journal}{Applied Physics Letters}}
  \textbf{\bibinfo{volume}{99}} (\bibinfo{year}{2011}).

\bibitem{zhang2024acceptor}
\bibinfo{author}{Zhang, Z.-H.} \emph{et~al.}
\newblock \bibinfo{title}{Acceptor-induced bulk dielectric loss in
  superconducting circuits on silicon}.
\newblock \emph{\bibinfo{journal}{arXiv preprint arXiv:2402.17155}}
  (\bibinfo{year}{2024}).

\bibitem{de2020two}
\bibinfo{author}{De~Graaf, S.} \emph{et~al.}
\newblock \bibinfo{title}{Two-level systems in superconducting quantum devices
  due to trapped quasiparticles}.
\newblock \emph{\bibinfo{journal}{Science Advances}}
  \textbf{\bibinfo{volume}{6}}, \bibinfo{pages}{eabc5055}
  (\bibinfo{year}{2020}).

\bibitem{wang2022towards}
\bibinfo{author}{Wang, C.} \emph{et~al.}
\newblock \bibinfo{title}{Towards practical quantum computers: Transmon qubit
  with a lifetime approaching 0.5 milliseconds}.
\newblock \emph{\bibinfo{journal}{npj Quantum Information}}
  \textbf{\bibinfo{volume}{8}}, \bibinfo{pages}{3} (\bibinfo{year}{2022}).

\bibitem{deng2023titanium}
\bibinfo{author}{Deng, H.} \emph{et~al.}
\newblock \bibinfo{title}{Titanium nitride film on sapphire substrate with low
  dielectric loss for superconducting qubits}.
\newblock \emph{\bibinfo{journal}{Physical Review Applied}}
  \textbf{\bibinfo{volume}{19}}, \bibinfo{pages}{024013}
  (\bibinfo{year}{2023}).

\bibitem{keller2017transmon}
\bibinfo{author}{Keller, A.~J.} \emph{et~al.}
\newblock \bibinfo{title}{Al transmon qubits on silicon-on-insulator for
  quantum device integration}.
\newblock \emph{\bibinfo{journal}{Applied Physics Letters}}
  \textbf{\bibinfo{volume}{111}} (\bibinfo{year}{2017}).

\bibitem{dixit2021searching}
\bibinfo{author}{Dixit, A.~V.} \emph{et~al.}
\newblock \bibinfo{title}{Searching for dark matter with a superconducting
  qubit}.
\newblock \emph{\bibinfo{journal}{Physical review letters}}
  \textbf{\bibinfo{volume}{126}}, \bibinfo{pages}{141302}
  (\bibinfo{year}{2021}).

\bibitem{clerk2010introduction}
\bibinfo{author}{Clerk, A.~A.}, \bibinfo{author}{Devoret, M.~H.},
  \bibinfo{author}{Girvin, S.~M.}, \bibinfo{author}{Marquardt, F.} \&
  \bibinfo{author}{Schoelkopf, R.~J.}
\newblock \bibinfo{title}{Introduction to quantum noise, measurement, and
  amplification}.
\newblock \emph{\bibinfo{journal}{Reviews of Modern Physics}}
  \textbf{\bibinfo{volume}{82}}, \bibinfo{pages}{1155--1208}
  (\bibinfo{year}{2010}).

\bibitem{burnett2013slow}
\bibinfo{author}{Burnett, J.} \emph{et~al.}
\newblock \bibinfo{title}{Slow noise processes in superconducting resonators}.
\newblock \emph{\bibinfo{journal}{Physical Review B—Condensed Matter and
  Materials Physics}} \textbf{\bibinfo{volume}{87}}, \bibinfo{pages}{140501}
  (\bibinfo{year}{2013}).

\bibitem{burnett2014evidence}
\bibinfo{author}{Burnett, J.} \emph{et~al.}
\newblock \bibinfo{title}{Evidence for interacting two-level systems from the
  1/f noise of a superconducting resonator}.
\newblock \emph{\bibinfo{journal}{Nature Communications}}
  \textbf{\bibinfo{volume}{5}}, \bibinfo{pages}{4119} (\bibinfo{year}{2014}).

\bibitem{burnett2019decoherence}
\bibinfo{author}{Burnett, J.~J.} \emph{et~al.}
\newblock \bibinfo{title}{Decoherence benchmarking of superconducting qubits}.
\newblock \emph{\bibinfo{journal}{npj Quantum Information}}
  \textbf{\bibinfo{volume}{5}}, \bibinfo{pages}{54} (\bibinfo{year}{2019}).

\bibitem{niepce2021stability}
\bibinfo{author}{Niepce, D.}, \bibinfo{author}{Burnett, J.~J.},
  \bibinfo{author}{Kudra, M.}, \bibinfo{author}{Cole, J.~H.} \&
  \bibinfo{author}{Bylander, J.}
\newblock \bibinfo{title}{Stability of superconducting resonators: Motional
  narrowing and the role of landau-zener driving of two-level defects}.
\newblock \emph{\bibinfo{journal}{Science advances}}
  \textbf{\bibinfo{volume}{7}}, \bibinfo{pages}{eabh0462}
  (\bibinfo{year}{2021}).

\bibitem{vallieres2024loss}
\bibinfo{author}{Valli{\`e}res, A.} \emph{et~al.}
\newblock \bibinfo{title}{Loss tangent fluctuations due to two-level systems in
  superconducting microwave resonators}.
\newblock \emph{\bibinfo{journal}{arXiv preprint arXiv:2412.05482}}
  (\bibinfo{year}{2024}).

\end{thebibliography}

\pagebreak

\begin{center}
\textbf{\large Supplemental Materials: Enhancing Intrinsic Quality Factors Approaching 10 Million in Superconducting Planar Resonators via Spiral Geometry}
\end{center}
\setcounter{section}{0}
\setcounter{equation}{0}
\setcounter{figure}{0}
\setcounter{table}{0}
\setcounter{page}{1}
\makeatletter
\renewcommand{\theequation}{S\arabic{equation}}
\renewcommand{\thefigure}{S\arabic{figure}}

\section{Line Calibration}\label{sec1}
To determine the power input to the resonators, we calibrated the cables and attenuators by measuring the S$_{21}$ transmission through the calibration line.
The calibration lines consist of the following four configurations:
\begin{enumerate}
  \item \textit{Direct-through line}: A direct connection between the input and output ports through the dilution refrigerator without any attenuators.
  \item \textit{46 dB attenuation line}: A connection with a built-in 46 dB attenuator inserted.
  \item \textit{Filtered attenuation line}:  A connection with an 8 GHz lowpass filter and a 20 dB attenuator inserted.
  \item \textit{Sample-equivalent line}: A configuration identical to the measurement sample, but switchable to a direct connection (as shown in Fig.\ \ref{fig0}(d)).
\end{enumerate}

These calibration lines allow us to separately evaluate the attenuation contributions from each component in the measurement system, including cables, attenuators, filters, and bonding losses. The evaluation was performed using the following steps:
\begin{enumerate}
  \item \textit{Cable loss evaluation}:  The transmission characteristics of the direct-through line were measured to determine the frequency-dependent attenuation of the cables running from room temperature to the inside of the dilution refrigerator.
  \item \textit{Attenuator evaluation}:  The attenuation of the 46 dB attenuation line was compared with the direct-through line to extract the actual frequency-dependent attenuation introduced by the attenuator.
  \item \textit{Filtered attenuator evaluation}:  The transmission characteristics of the filtered attenuation line were compared with those of the direct-through line to obtain the attenuation characteristics of the lowpass filter with an additional 20 dB attenuator.
  \item \textit{Bonding loss evaluation}:  The sample-equivalent line was compared with the measured sample to evaluate the reflection loss caused by bonding between the chip and the substrate.
\end{enumerate}

\begin{figure}
  \centering
  \includegraphics[width=\linewidth]{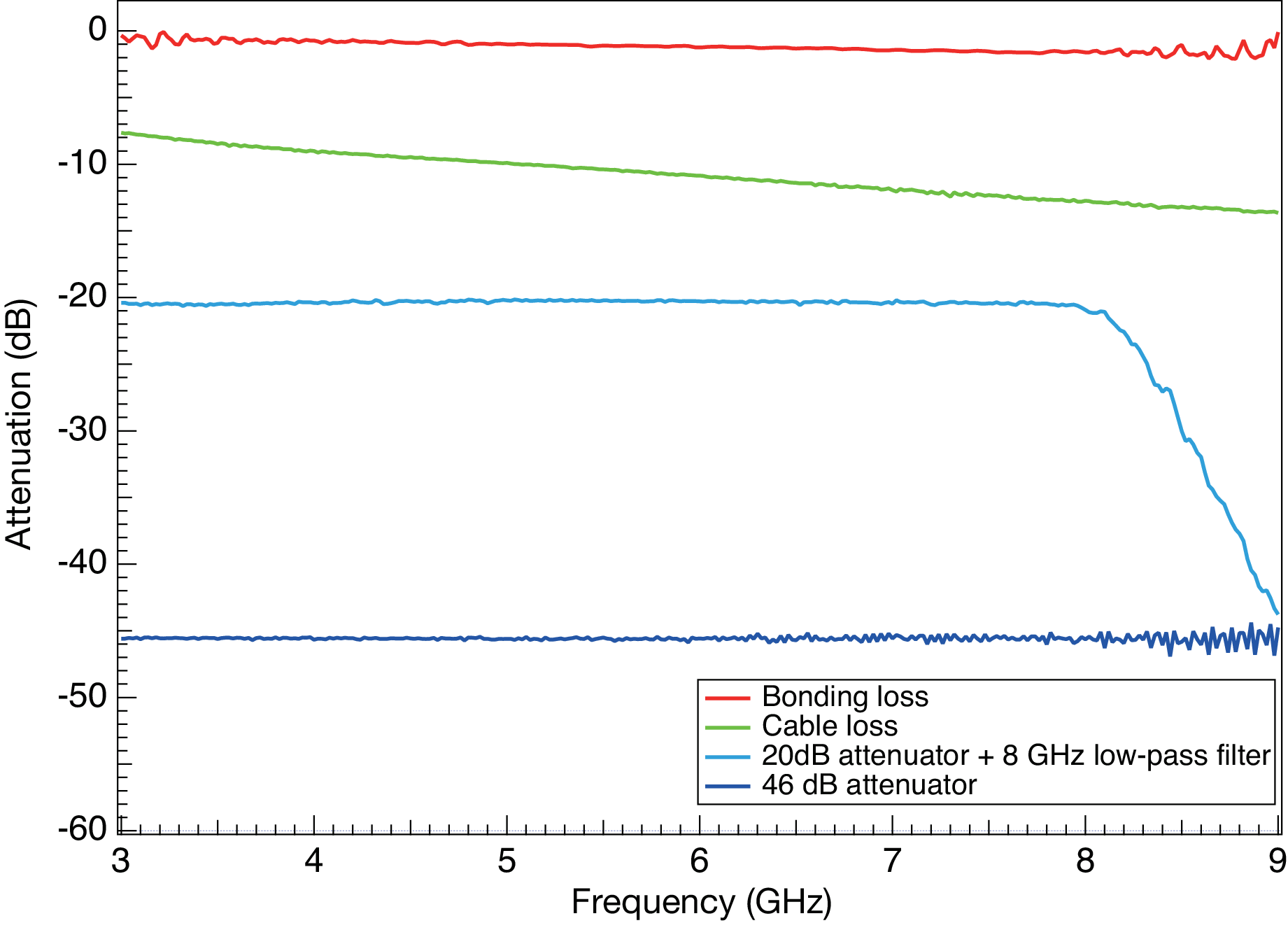}
  \caption{Measured attenuation characteristics of different components, including bonding loss, cable loss, attenuators, and filters, as a function of frequency. These losses were evaluated by comparing the transmission of calibration lines with and without each component.}
  \label{fig1s1}
\end{figure}

Based on these measurements, the frequency-dependent attenuation values were obtained for each contribution as shown in Fig.\ \ref{fig1s1}. The sum of these values determined the total input power attenuation to the resonator. At resonator frequencies of 4.02, 4.49, 4.81, 5.34, 6.01, 6.16, and 6.86 GHz, the total attenuation values were -75.8 dB, -76.1 dB, -76.5 dB, -77.0 dB, -77.9 dB, -78.1 dB, and -79.0 dB, respectively.

The photon number $\bar{n}$ inside the resonator can be calculated using the input power $P_\mathrm{in}$ accounting for this attenuation and considering the external quality factor $Q_\mathrm{c}$ and internal quality factor $Q_\mathrm{i}$ as follows \cite{clerk2010introduction}:
\begin{align}
  \bar{n} = \frac{4}{2\pi f Q_\mathrm{c}} \lr{\frac{1}{Q_\mathrm{c}}+ \frac{1}{Q_\mathrm{i}}}^{-2} \frac{P_\mathrm{in}}{h f}.
\end{align}

\section{Loss Tangent Fluctuations}\label{sec2}

At low excitation power, the internal loss rate fluctuates at low frequencies, leading to multiple quasi-stable loss rates. These fluctuations are attributed to two-level systems in the dielectric materials, which exhibit different configurations at different time \cite{burnett2013slow, burnett2014evidence, burnett2019decoherence, niepce2021stability}.

To investigate this behavior, we measured the S$_{21}$ transmission of ASR1 and CPW1 over a 24-hour period. The output power from the VNA was set to $-$80 dBm, with an additional 6 dB attenuator at room temperature. This setup corresponded to average photon numbers of 4.8 for ASR1 and 1.9 for CPW1. The intermediate frequency (IF) bandwidth of the VNA was set to 5 Hz which allowed for quick measurements before significant fluctuations occurred, while keeping the noise level sufficiently low to enable accurate measurement of the quality factors of the resonators.

\begin{figure}[b]
  \includegraphics[width = \linewidth]{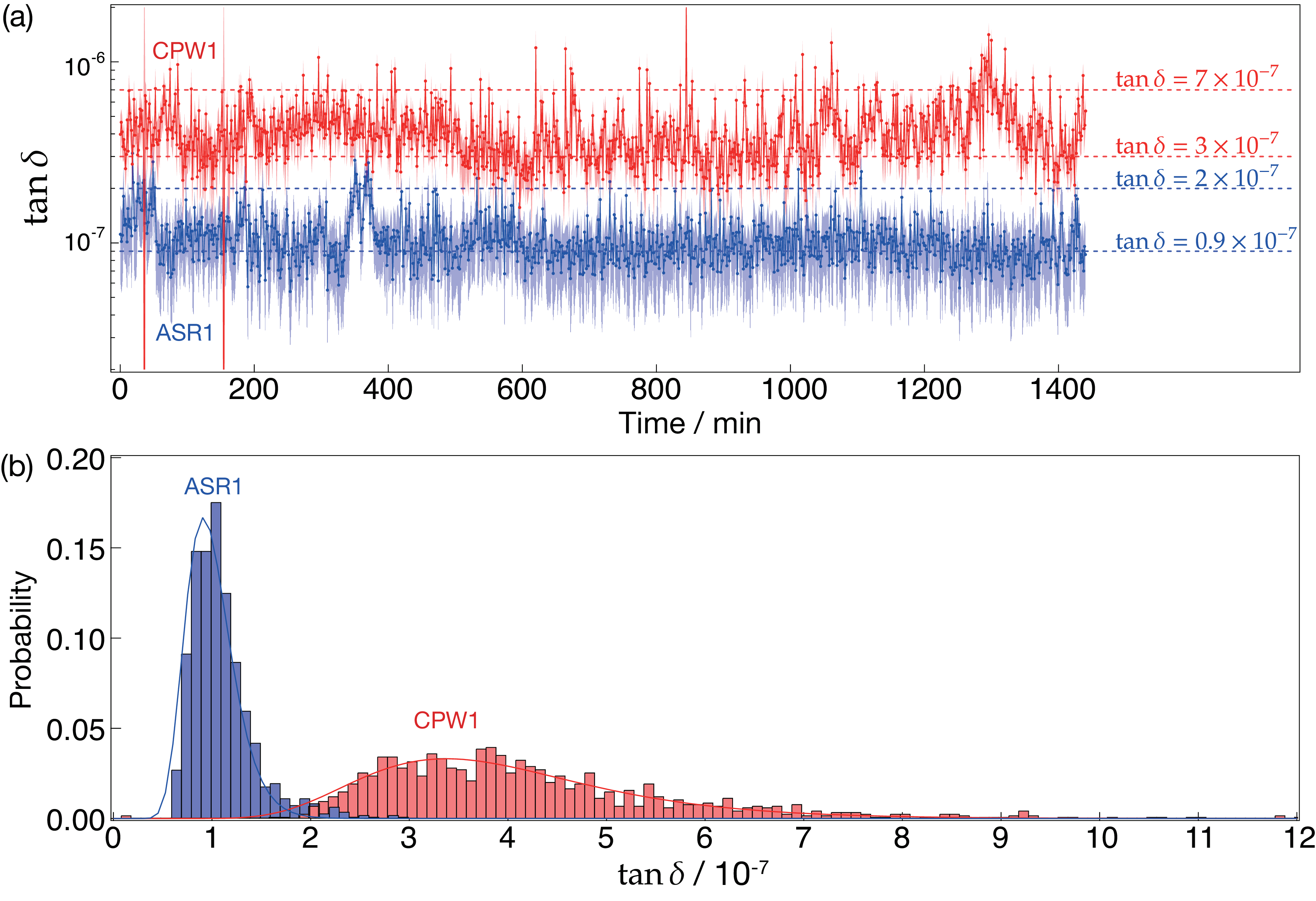}
  \caption{(a) Time-series data of the loss tangent for ASR1 and CPW1, with distinct jumps in the loss rate indicating transitions between quasi-stable loss states. These fluctuations are attributed to the collective switching of two-level systems (TLS) in the dielectric materials. (b) The histograms of the loss tangent values extracted from (a). Both histograms exhibit long tails toward higher loss values, which are fitted to log-normal distributions with parameters $x_0 = 9.11 \times 10^{-8}$ and $\sigma = 0.42$ for ASR1 and $x_0 = 3.37 \times 10^{-7}$ and $\sigma = 0.48$ for CPW1.}
  \label{fig_s2}
\end{figure}

The acquired data exhibit distinct jumps in the loss rate over time, as illustrated in Fig.\ \ref{fig_s2}(a). Periods of relatively stable loss are intermittently interrupted by abrupt transitions, indicating shifts between different loss states. This characteristic behavior suggests a mechanism governed by the collective switching of ensembles of two-level systems (TLS) present in the dielectric materials.

Interestingly, these fluctuations appear to be significantly more pronounced in CPW1 compared to ASR1. This disparity likely arises from the differing electric field distributions in the two geometries, which can influence the interaction between the TLS and the electromagnetic environment. In particular, CPW structures tend to exhibit stronger field concentrations near material interfaces, where TLS defects are more abundant, leading to enhanced loss fluctuations.

To further quantify these variations, Fig.\ \ref{fig_s2}(b) presents histograms of the loss tangent extracted from the time-series data in Fig.\ \ref{fig_s2}(a). These histograms reveal a broad distribution with extended tails toward higher loss values, indicating the presence of rare but significant loss events. Such distributions are well-described by log-normal statistics, which are commonly observed in systems exhibiting multiplicative noise \cite{vallieres2024loss}. The probability density function of the log-normal fit is given by
\begin{align}
  f(x) \propto \explr{- \lr{\frac{\ln (x/x_0)}{2\sigma^2}}^2}
\end{align}
where the fit parameters are $x_0 = 9.11 \times 10^{-8}$ and $\sigma = 0.42$ for ASR1 and $x_0 = 3.37 \times 10^{-7}$ and $\sigma = 0.48$ for CPW1. These values indicate that, on average, CPW1 exhibits higher loss than ASR1 and a broader spread in loss values, reinforcing the notion that TLS-induced noise is more significant in CPW structures.

\section{Voltage Distribution in Spiral Resonators}\label{sec3}
\begin{figure}
  \centering
  \includegraphics[width=0.6\linewidth]{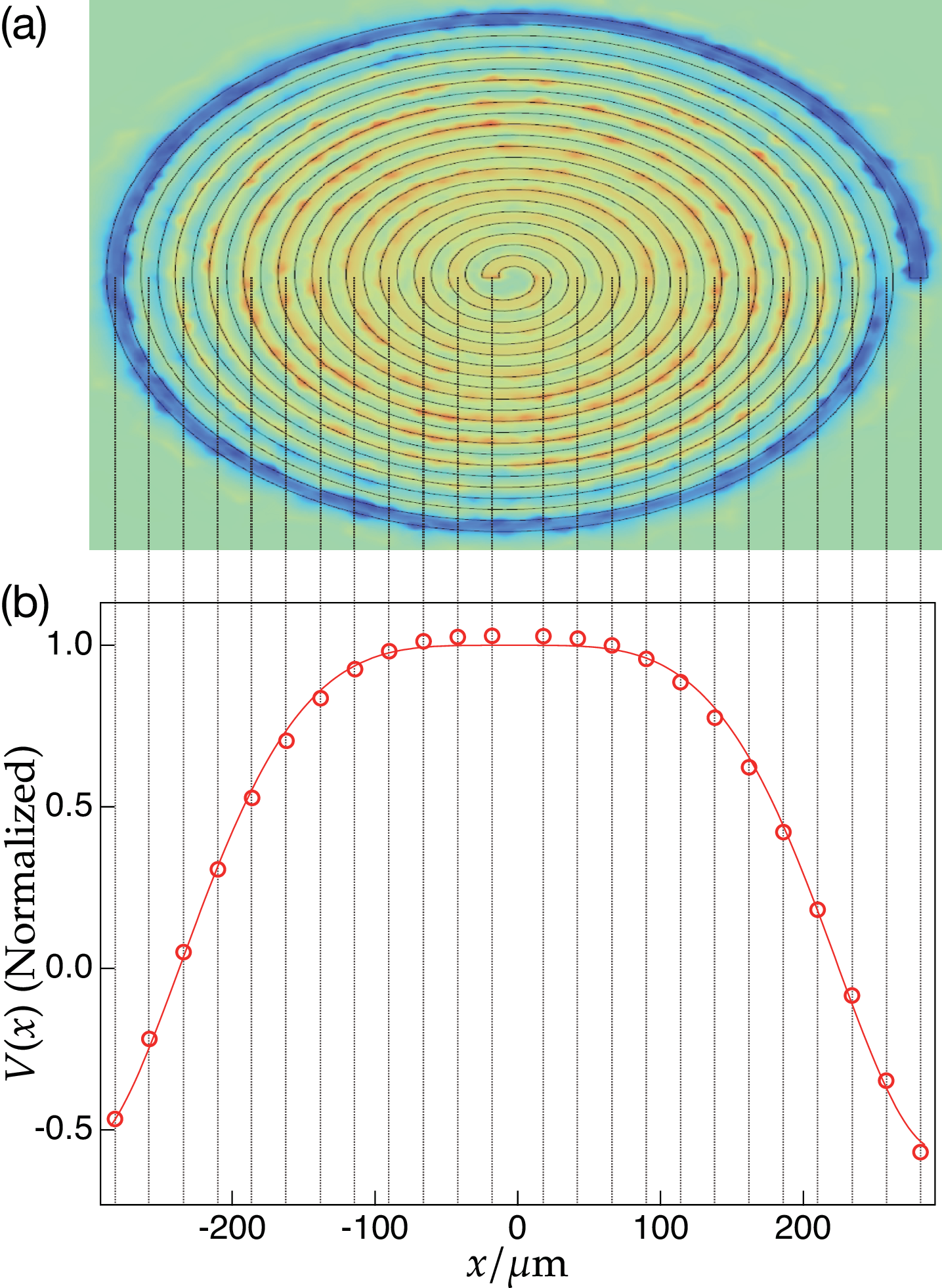}
  \caption{The COMSOL simulation shows the electric field distribution (a) and normalized voltage distribution (b) for an ASR with 12 turns, 24 \textmu m pitch, and 12 \textmu m gap. The simulated points closely match the analytical model. The voltage distribution at each point was calculated by integrating the electric field from the ground plane to each ASR arm.}
  \label{figs3}
\end{figure}

To verify the validity of the theoretical equation Eq.\ \ref{eq13} for the electric field distribution in the ASR, we conducted simulations using COMSOL. A three-dimensional model of an ASR was constructed with number of spiral turns at 12 and pitch of 12 \textmu m with gaps 12 \textmu m, and an eigenfrequency analysis was performed to determine the voltage distribution. The results are presented in Fig.\ \ref{figs3}. The simulation results align well with the theoretical predictions, demonstrating the accuracy of the theoretical model.

\end{document}